\documentstyle[12pt]{article}
\begin{document}
\topmargin=-1.0cm
\evensidemargin=0cm
\oddsidemargin=0cm
\newcommand{\BQ}{\begin{equation}}
\newcommand{\EQ}{\end{equation}}
\newcommand{\BQA}{\begin{eqnarray}}
\newcommand{\EQA}{\end{eqnarray}}
\newcommand{\half}{\frac{1}{2}}
\newcommand{\NN}{\nonumber \\}
\newcommand{\E}{{\rm e}}
\newcommand{\del}{\partial}
\renewcommand{\thefootnote}{\fnsymbol{footnote}}
\newcommand{\zm}[1]{\stackrel{\circ} {#1} }
\newcommand{\nzm}[1]{\widetilde {#1} }
\newcommand{\llangle}{\langle \langle}
\newcommand{\rrangle}{\rangle \rangle}
\date{}
\baselineskip=0.6cm
\renewcommand{\appendix}{\renewcommand{\thesection}
{\Alph{section}}\renewcommand{\theequation}{\Alph{section}.\arabic{equation}}
\setcounter{equation}{0}\setcounter{section}{0}}
\begin{titlepage}
\begin{flushright}
{KOBE-TH-99-03\\ IFUP-TH 7/99}
\end{flushright}
\vspace{1cm}
\begin{center}
{\LARGE Spontaneous Supersymmetry 
Breaking}\\
\vspace{5mm}
{\LARGE from Extra Dimensions }\\
\vskip1.0truein
{\large Makoto Sakamoto}$^{(a)}$,
\footnote{E-mail: {\tt sakamoto@oct.phys.kobe-u.ac.jp}},
{\large Motoi Tachibana}$^{(b)}$ and 
\footnote{E-mail: {\tt tatibana@oct.phys.kobe-u.ac.jp}}
{\large Kazunori Takenaga}$^{(c)}$ 
\footnote{E-mail: {\tt takenaga@ibmth.df.unipi.it}} 
\vskip0.2truein
\centerline{$^{(a)}$ {\it Department of Physics,
Kobe University, Rokkodai, Nada, Kobe 657-8501, Japan}}
\vspace*{2mm}
\centerline{$^{(b)}$ {\it Graduate School of Science and Technology,
Kobe University, Rokkodai, Nada, Kobe 657-8501, Japan }}
\vspace*{2mm}
\centerline{$^{(c)}$ {\it 
I.N.F.N, Sezione di Pisa, Via Buonarroti, 2 Ed. B,
56127 Pisa, Italy}}
\end{center}
\vskip0.5truein \centerline{\bf Abstract} \vskip0.13truein
We propose a new spontaneous supersymmetry breaking mechanism,
in which extra compact dimensions play an important role.
To illustrate our mechanism, we study a simple model consisting of two
chiral superfields, where one spatial dimension is compactified
on a circle $S^1$. It is shown that supersymmetry is spontaneously
broken irrespective of the radius of the circle, and also that
the translational invariance for the $S^1$-direction 
and a global symmetry are
spontaneously broken when the radius becomes larger than a
critical radius. These results are expected to be general features 
of our mechanism. We further discuss that our mechanism may be
observed as the O'Raifeartaigh type of supersymmetry breaking
at low energies.
\end{titlepage}
\newpage
\baselineskip 20 pt
\pagestyle{plain}
\vskip0.2truein
\setcounter{equation}{0}
\vskip0.2truein
\section{Introduction}
Recently, a very challenging possibility of large-scale
compactification has been pointed out in ref.\cite{LSC}.
The authors have discussed consequences of extra large-scale
dimensions and have proposed phenomenologically interesting
scenarios. So, an extremely exciting situation will be that not
only extra dimensions but also supersymmetry might be observed
in the near future. If the scale of extra dimensions is not far from
that of supersymmetry breaking, it will be natural to think that
supersymmetry breaking and compactification of extra dimensions
may have a common dynamical origin.
\par
The purpose of this paper is to propose a new spontaneous 
supersymmetry breaking mechanism by compactification. Let us
first present a key idea of our mechanism below. Suppose that
an (effective) potential $V(A_i)$ for scalar fields $A_i$ vanishes
at some values $\bar{A_i }$ of $A_i$, i.e.
\BQ
V(A_i)\Big|_{A_i = \bar{A_i }} = 0.
\label{pot}
\EQ
One might then conclude that supersymmetry would be unbroken
because of the vanishing vacuum energy. However, it is {\it not}
always true. This is an essential point in our mechanism. Our
spontaneous supersymmetry breaking mechanism will be a realization
of the following simple idea: If there exist some mechanisms to
force vacuum expectation values of $A_j$ for some $j$ not to take
the values $\bar{A_j}$, then the vanishing vacuum energy solution
$\bar{A_i}$ in eq.(\ref{pot}) will not be realized as a supersymmetric
vacuum
\footnote{An interesting example of such mechanism has been found
in a special class of supersymmetric models \cite{YANA}, in which
would-be supersymmetric vacuum configurations have been removed
from quantum moduli spaces due to quantum deformed constraints.}.
A simple mechanism to force vacuum expectation values not to take
nonzero constants has been proposed in ref.\cite{SAKA}. We shall
apply this mechanism for supersymmetric field theories to break
supersymmetry spontaneously.
\par
In the next section, to illustrate our mechanism, we shall study a 
3+1-dimensional Wess-Zumino type model in which one spatial
dimension is compactified on a circle $S^1$ and show that supersymmetry
is spontaneously broken. In Sect.3, it is shown that the translational
invariance for the $S^1$-direction and a global symmetry are spontaneously
broken when the radius of the circle becomes larger than a critical
radius. In Sect.4, our mechanism is contrasted with the O'Raifeartaigh
mechanism \cite{OR}. Some comments are given in the last section.
\section{A Model}
To illustrate our spontaneous supersymmetry breaking mechanism,
let us consider a 3+1-dimensional Wess-Zumino type model consisting
of two chiral superfields $\Phi_0$ and $\Phi_1$. The superpotential
we take is 
\BQ
W(\Phi_0,\Phi_1) = g\Phi_0\left( \frac{\Lambda^2}{g^2}
- \half(\Phi_1)^2 \right),
\label{spot}
\EQ
where the parameters $g$ and $\Lambda$ are chosen to be real
and positive for simplicity. This model has a global $Z_2$ symmetry
\BQA
\Phi_0 &\longrightarrow& +\Phi_0 , \NN
\Phi_1 &\longrightarrow& -\Phi_1.
\label{Z2}
\EQA
It turns out that this global symmetry plays an important role in
this model. The scalar potential is given by
\BQ
V(A_0,A_1) = |F_0|^2 +|F_1|^2, 
\label{spot2}
\EQ
where $A_0$ and $A_1$ denote the lowest scalar components of 
$\Phi_0$ and $\Phi_1$, respectively, and 
\BQA
F_0 &=&  -\left( \frac{\del W(A_0,A_1)}{\del A_0} \right)^*
= - \frac{\Lambda^2}{g} + \frac{g}{2}(A^*_1)^2, \NN
F_1 &=& -\left( \frac{\del W(A_0,A_1)}{\del A_1}\right)^*
= g A^*_0 A^*_1.
\label{fcomp}
\EQA
Since the scalar potential $V(A_0,A_1)$ would vanish at $A_0$ = 0
and $A_1$ = $\pm{\frac{\sqrt{2}\Lambda}{g}}$, supersymmetry
might be unbroken, while the $Z_2$ symmetry be broken, spontaneously.
This is, however, a hasty conclusion, as we will see below.
\par
Let us suppose that one of the space coordinates, say, $y \equiv x^3$
is compactified on a circle $S^1$ whose radius is $R$. Since $S^1$
is multiply-connected and the action has the $Z_2$ symmetry
(\ref{Z2}), we can impose the following nontrivial boundary conditions
associated with the $Z_2$ symmetry:
\BQA
\Phi_0 (x^{\mu}, y+2\pi R) &=&  +\Phi_0 (x^{\mu}, y), \NN  
\Phi_1 (x^{\mu}, y+2\pi R) &=&  -\Phi_1 (x^{\mu}, y),  
\label{bc}
\EQA
where $x^{\mu}$ denote the coordinates of
the uncompactified 2+1 -dimensional
Minkowski spacetime. It should be stressed that the boundary
conditions (\ref{bc}) are consistent with supersymmetry and that
the action is still single-valued thanks to the $Z_2$ symmetry
(\ref{Z2}). An important consequence of the nontrivial boundary
conditions (\ref{bc}) is that any vacuum expectation value of
$\Phi_1(x^{\mu},y)$ (or $A_1(x^{\mu},y)$) cannot be a 
($y$-independent) nonzero constant. It immediately follows that
(would-be) supersymmetric vacuum configurations $A_0 = 0$ and
$A_1 = \pm{\frac{\sqrt{2}\Lambda}{g}}$ should be ruled out.
If we assume that the vacuum would translationally be invariant
\footnote{This assumption is true only when the radius of the circle
is smaller than a critical radius $R^*$ = $\frac{1}{2\Lambda}$.
See the next section.},
the vacuum expectation value of $A_1(x^{\mu},y)$ has to vanish,
i.e.
\BQ
\langle A_1(x^{\mu},y) \rangle = 0.
\label{vev}
\EQ
Replacing $A_0$ and $A_1$ by their vacuum expectation values
in $V(A_0,A_1)$, we find 
\BQ
V(\langle A_0 \rangle, \langle A_1 \rangle = 0) = 
\frac{\Lambda^4}{g^2} > 0,
\label{vacuum}
\EQ
which implies that supersymmetry is spontaneously broken, as expected.
In this model, there is a flat direction in vacuum configurations 
since the potential (\ref{vacuum}) is independent of 
$\langle A_0 \rangle$.
\par
Another way to see the supersymmetry breaking more explicitly
may be to expand the component fields in the Fourier-series according to
the boundary conditions (\ref{bc}).
\BQA
A_0(x^{\mu}, y) &=& \frac{1}{\sqrt{2\pi R}}\sum^{\infty}_{n=-\infty}
a^{(2n)}_0(x^{\mu})\ e^{i2n\frac{y}{2R}}\ ,  \NN  
A_1(x^{\mu}, y) &=& \frac{1}{\sqrt{2\pi R}}\sum^{\infty}_{l=-\infty}
a^{(2l-1)}_1(x^{\mu})\ e^{i(2l-1)\frac{y}{2R}}\ ,  \NN  
\psi_0(x^{\mu}, y) &=& \frac{1}{\sqrt{2\pi R}}\sum^{\infty}_{n=-\infty}
\chi^{(2n)}_0(x^{\mu})\ e^{i2n\frac{y}{2R}}\ ,  \NN  
\psi_1(x^{\mu}, y) &=& \frac{1}{\sqrt{2\pi R}}\sum^{\infty}_{l=-\infty}
\chi^{(2l-1)}_1(x^{\mu})\ e^{i(2l-1)\frac{y}{2R}}\ . 
\label{mode}
\EQA
It turns out to be convenient to divide the Fourier mode $a^{(2l-1)}_1$
into two parts as
\BQ
a^{(2l-1)}_1 = a^{(2l-1)}_- + i a^{(2l-1)}_+
\label{a1}
\EQ
with $a^{(2l-1)*}_{\pm}$ = $a^{(-2l+1)}_{\pm}$. Then the squared
masses for $a^{(2n)}_0$, $\chi^{(2n)}_0$, 
$a^{(2l-1)}_{\pm}$ and $\chi^{(2l-1)}_1$ (in a viewpoint of the 
2+1-dimensional
Minkowski spacetime) are given by $m^2$ =
$\left(\frac{n}{R}\right)^2$, $\left(\frac{n}{R}\right)^2 $, 
$\pm \Lambda^2  + |M|^2$ + $\left( \frac{l-\half}{R}\right)^2$,
$|M|^2$ + $\left( \frac{l-\half}{R}\right)^2$,
respectively, where $M$ = $g\langle A_0 \rangle$. Here we would
like to make several comments on the mass spectrum. The first
comment is that the supersymmetry breaking scale is found, from
the mass splitting, to be of the order of $\Lambda$. The second comment is
that the mass spectrum satisfies the following relations
\cite{FER}:
\BQA
m^2_{a^{(2n)}_0} &=&  m^2_{\chi^{(2n)}_0}, \NN
m^2_{a^{(2l-1)}_-}+m^2_{a^{(2l-1)}_+} &=& 2m^2_{\chi^{(2l-1)}_1}.
\label{a1mass}
\EQA
The third comment is that the fermionic mode $\chi^{(0)}_0$ is
massless and corresponds to the Nambu-Goldstone fermion associated
with the spontaneous supersymmetry breaking. The bosonic partner
$a^{(0)}_0$ is also massless but its origin is quite different. A part
of it will correspond to the Nambu-Goldstone boson associated with
the spontaneous breaking of a $U(1)_R$ symmetry (with $\langle A_0 \rangle
\neq 0$). The masslessness of $a^{(0)}_0$ is also guaranteed
by the existence of a flat direction of $\langle A_0 \rangle$ (at least at
the tree level). The last comment is that choosing $\langle A_0 \rangle  =
0$ we find that some of the bosonic modes  $a^{(2l-1)}_-$ might have
negative squared masses for $R > R^* = \frac{1}{2\Lambda}$. This
observation suggests that the configuration (\ref{vev}) may become
unstable for $R > R^*$ and that a phase transition can occur at $R = R^*$.
This is the subject of the next section. 
\section{Spontaneous Breakdown of Translational Invariance}
In the previous section, we have assumed 
the translational invariance would
be unbroken. It turns out that this assumption is not true for $R > R^*$,
as suggested in the previous section. 
We shall here discuss spontaneous breakdown of 
the translational invariance for the $S^1$-direction. 
To this end, we should take account of kinetic terms 
as well as potential terms
since the vacuum configuration might be coordinate-dependent. 
The vacuum configuration will then be obtained 
by solving a minimization problem of the functional
\footnote{The ${\cal E}[A_0,A_1]$ may be thought 
of as a potential in a viewpoint of the 2+1-dimensional 
Minkowski spacetime.}
\BQ
{\cal E}[A_0,A_1;R] = \int^{2\pi R}_0 dy \left\{
\bigg|\frac{\del A_0}{\del y}\bigg|^2 + 
\bigg|\frac{\del A_1}{\del y}\bigg|^2
+ V(A_0,A_1) \right\},
\label{energy}
\EQ
with the boundary conditions
\BQA
A_0(y+2\pi R) &=& +A_0(y), \NN
A_1(y+2\pi R) &=& -A_1(y). 
\label{bc2}
\EQA
In the following, we ignore the $x^{\mu}$ dependence 
since we are interested in the vacuum configuration, 
for which the translational invariance of the 2+1-dimensional 
Minkowski spacetime is assumed to be unbroken. 
We first note that the vacuum configuration for $A_0(y)$
and $A_1(y)$ should satisfy the following field equations:
\BQA
0 &=&\frac{\delta {\cal E}[A_0,A_1;R]}{\delta A^*_0(y)} =
-\frac{d^2 A_0(y)}{dy^2}+g^2 A_0(y)|A_1(y)|^2, \NN
0 &=&\frac{\delta {\cal E}[A_0,A_1;R]}{\delta A^*_1(y)} =
-\frac{d^2 A_1(y)}{dy^2}-\Lambda^2 A^*_1(y) \NN
& & \qquad \qquad  \qquad  \qquad  \qquad  
+ g^2\left( |A_0(y)|^2+\half|A_1(y)|^2\right)A_1(y).
\label{fieldeq}
\EQA
If the translational invariance for the $S^1$-direction would be unbroken, 
the vacuum expectation value of $A_1$ has to vanish due to 
the boundary conditions
(\ref{bc2}) and then the functional ${\cal E}[A_0,A_1;R]$ becomes
\BQ
{\cal E}[A_0 = {\rm const},A_1=0;R] = \frac{2\pi R \Lambda^4}{g^2}.
\label{energy2}
\EQ
Using the field equations (\ref{fieldeq}) to eliminate the \lq \lq 
kinetic'' terms in eq.(\ref{energy}), we may find 
\BQA
{\cal E}[A_0 ,A_1;R]\Big|_{\frac{\delta{\cal E}}{\delta A_0}
= \frac{\delta{\cal E}}{\delta A_1}=0} 
&=& \frac{2\pi R \Lambda^4}{g^2}-\int^{2\pi R}_0 dy\left\{
\frac{g^2}{4}|A_1|^4+g^2|A_0A_1|^2 \right\} \NN
&\leq& {\cal E}[A_0 = {\rm const},A_1=0;R].
\label{energy3}
\EQA
We have thus arrived at an important conclusion: If there would appear
nontrivial solutions ($A_1 \neq 0$) to the field equations (\ref{fieldeq}),
then $A_1$ = 0 is no longer a vacuum configuration and 
the translational invariance for the $S^1$-direction 
would then be broken spontaneously (with the
$Z_2$ symmetry breaking) since nonvanishing $A_1(y)$ inevitably
has the $y$ dependence to be consistent with the boundary conditions
(\ref{bc2}). We should again emphasize that (would-be) $y$-independent
solutions $A_0 = 0$ and $A_1 = \pm\frac{\sqrt{2}\Lambda}{g}$ to
eqs.(\ref{fieldeq}) are not consistent with eqs.(\ref{bc2}). 
It turns out that for $R \leq R^* = \frac{1}{2\Lambda}$ 
there exists only the trivial solution ($A_0$ = const. and $A_1$ = 0) 
to eqs.(\ref{fieldeq}), 
while for $R> R^*$
there will appear many other (nontrivial) solutions. 
This result may be seen by noting that a vacuum configuration
with $A_{1}\ne0$ can be realized only when $A_{0}=0$ and
Im$A_{1}=0$, and then by solving eqs.(\ref{fieldeq})
with the boundary conditions (\ref{bc2}).
The vacuum configuration for $A_0$ and $A_1$, 
which minimizes ${\cal E}[A_0 ,A_1;R]$, 
has finally been found to be \cite{SAKA2}
%
%
\BQA
\langle A_0(x^{\mu},y) \rangle 
&=& { \,{\rm arbitrary \ constant}   \qquad \ {\rm for} \quad R \leq R^*
\atopwithdelims\{. \,\,0 \qquad \qquad 
\qquad  \qquad \qquad {\rm for }\quad R > R^*,} \NN
\langle A_1(x^{\mu},y) \rangle 
&=& { 0   \qquad \qquad \qquad \qquad \qquad {\rm for} \quad R \leq R^*
\atopwithdelims\{. \,\frac{2k\omega}{g}{\rm sn}\left(\omega(y-y_0),k
\right) \qquad  {\rm for }\quad R > R^*,} 
\label{vev2}
\EQA
with $\omega = \frac{\Lambda}{\sqrt{1+k^2}}$. Here, ${\rm sn}(u,k)$ is the
Jacobi elliptic function whose period is $4K(k)$, where $K(k)$ denotes
the complete elliptic function of the first kind. The parameter $k$
($0 \leq k < 1$) and the radius $R$ should be related through the 
equation
\BQ
R = \frac{K(k)}{\pi \omega}.
\label{radius}
\EQ
Note that as $k$ runs from zero to one the right hand side of 
eq.(\ref{radius}) increases monotonically from $R^* = \frac{1}{2\Lambda}$ 
to infinity. 
As expected in the previous section, a phase transition occurs at $R = R^*$ 
and the translational invariance for the $S^1$-direction
is spontaneously broken for $R > R^*$.
\par
We would like to comment on the normal modes of oscillation about 
the vacuum configuration for $R > R^*$. In the previous section, 
we have observed that some of Fourier modes might have negative 
squared masses for $R > R^*$. This is merely due to the fact that 
we have not taken the true vacuum configuration for $R > R^*$. 
In fact, we can show that all squared masses are positive 
semi-definite, as they should be, if the fields are correctly 
expanded in the normal modes of oscillation about the true 
vacuum configuration (\ref{vev2})
\footnote{We have not, however, found any simple relations 
between bosonic and fermionic masses, such as eqs.(\ref{a1mass}), 
though we can formally prove the vanishing of the supertrace 
${\rm Str} {} m^2$ in the whole space of the spectrum.}.
We may then find three massless modes 
(in a sense of real degrees of freedom):
One is bosonic and two are fermionic. 
The massless fermionic (bosonic) modes correspond to 
the Nambu-Goldstone modes associated with spontaneous 
breakdown of supersymmetry 
(the translational invariance for the $S^1$-direction). 
\section{O'Raifeartaigh Mechanism vs. Ours}
In this section, we would like to point out that our mechanism may be 
observed as the O'Raifeartaigh type of supersymmetry breaking 
at low energies, even though we will also point out several 
differences between the O'Raifeartaigh mechanism and ours.
\par
We shall first summarize general settings to construct 
supersymmetric models based on our mechanism. Let $\bar{A_i}$ 
be a (would-be) supersymmetric vacuum configuration satisfying 
$V(\bar{A_i}) = 0$
or $\frac{\del W(\bar{A_i})}{\del A_j} = 0$ for all $j$. 
Suppose that some of space dimensions are compactified 
on a manifold which should be 
translationally invariant and be 
multiply-connected, like $S^1$. 
We then impose nontrivial boundary conditions on superfields, 
which have to be consistent with global symmetries of the theory. 
The crucial point is that the boundary conditions have to be chosen 
to prevent some of vacuum expectation values of $A_i$ 
from taking the values $\bar{A_i}$. 
It turns out \cite{SAKA2} that at least {\it two} chiral superfields 
are required for our mechanism to work in Wess-Zumino type models 
and that the model
presented in Sect. 2 is the minimum one. This may be contrasted with the 
O'Raifeartaigh mechanism, in which at least {\it three} chiral superfields 
are required. In the O'Raifeartaigh mechanism, superpotentials $W(A_i)$ 
should be chosen such that there are no consistent solutions to the equations
\BQ
\frac{\del W(A_i)}{\del A_j} = 0  \qquad {\rm for \  all} \  j.
\label{or}
\EQ
On the other hand, in our mechanism superpotentials will be chosen 
to have (would-be) solutions to eqs.(\ref{or}) but to have no solutions if
we further impose boundary conditions on $A_i$, which have to be 
inconsistent with eqs.(\ref{or}). In this point, our mechanism is 
apparently different from the O'Raifeartaigh one.
\par
Let us next discuss a resemblance between the two mechanisms. 
To make our discussions simple, we will consider the model studied 
in Sect.2 again. Let $W(a^{(2n)}_0, a^{(2l-1)}_1)$ be the superpotential 
for the Fourier modes given in eq.(\ref{mode}). Since the mode expansions 
(\ref{mode}) have been done in a consistent way with the boundary 
conditions (\ref{bc2}), we may not need to take care of boundary 
conditions any more, as long as the Fourier modes $a^{(2n)}_0$ and
$a^{(2l-1)}_1$ are considered. Then, supersymmetry breaking might 
be observed by showing that the equations
\BQ
\frac{\del W(a^{(2n)}_0,a^{(2l-1)}_1)}{\del a^{(2m)}_0}
= \frac{\del W(a^{(2n)}_0,a^{(2l-1)}_1)}{\del a^{(2k-1)}_1} = 0
\qquad {\rm for \  all} \ m \ {\rm and} \ k
\label{eqm}
\EQ
have no consistent solutions. 
In this sense, our mechanism might be thought of as a kind
 of the O'Raifeartaigh one, though the model consists
of infinitely many (Kaluza-Klein) modes. 
To see the resemblance between two mechanisms 
more explicitly, let us look at the model from a low energy point of view.
To this end, we shall restrict our considerations to the light 
five bosonic modes, $a^{(0)}_0$,$a^{(\pm 2)}_0$ and $a^{(\pm 1)}_1$
\footnote{For $R > R^*$, the fields should appropriately be expanded 
in the normal modes of oscillation about the true vacuum 
configuration (\ref{vev}).},
and simply put other \lq\lq heavy" modes to be zero.
We may then find the equations (\ref{eqm}) for 
$a^{(0)}_0$,$a^{(\pm 2)}_0$ and $a^{(\pm 1)}_0$ to be
\BQA
a^{(1)}_1 a^{(-1)}_1 &=& \frac{2\pi R\Lambda^2}{g^2}, \NN
(a^{(\mp 1)}_1)^2  &=& 0, \NN
a^{(0)}_0 a^{(\mp 1)}_1 + a^{(\mp 2)}_0 a^{(\pm 1)}_1
&=& 0,
\label{eqm2}
\EQA
respectively. It is easy to see that these equations have no 
consistent solutions. This observation suggests that our 
mechanism may be observed as the O'Raifeartaigh type of 
supersymmetry breaking at low
energies.
\section{Comments}
We have studied a simple 3+1-dimensional supersymmetric model, in 
which one spatial dimension 
is compactified on a circle, to illustrate our spontaneous 
supersymmetry breaking mechanism. It has been shown that 
supersymmetry is spontaneously broken for $R > 0$ and also 
that the translational invariance for the $S^1$-direction 
with the global $Z_2$ 
symmetry is spontaneously broken for $R > R^*$. 
These results will not be specific to this model 
but are expected to be general features of our mechanism.
\par
We should make a comment on the Scherk-Schwarz mechanism 
\cite{SCH}. 
One might impose nontrivial boundary conditions associated with 
a $U(1)_R$ symmetry. Then, bosonic components of superfields 
may satisfy different boundary conditions from fermionic ones. 
A crucial difference between the Scherk-Schwarz mechanism 
and ours is that the breaking \`a la Scherk-Schwarz is {\it explicit}
rather than spontaneous at the level of global supersymmetry. 
Another difference is that the Scherk-Schwarz mechanism 
will work for any choice of superpotentials, just like 
supersymmetry breaking at finite temperature, 
while our mechanism will not.
\par
The final comment is as follows: 
An interesting supersymmetry breaking mechanism 
by compactification has been proposed by Dvali and Shifman \cite{DVALI}, 
who have called it dynamical compactification. The authors have 
suggested the idea that our Universe could spontaneously be 
generated in the form of a four-dimensional topological or 
non-topological stable defect in higher-dimensional spacetime 
and that the low-energy observers trapped in the core of 
the defect would not detect supersymmetry, although 
the vacuum of the original higher-dimensional 
theory is fully supersymmetric. What we would like to point out 
is that the supersymmetric model presented in Sect.2 may be 
thought of as an explicit realization of dynamical compactification 
in the limit of $R \rightarrow \infty$ ($k \rightarrow 1)$. 
In this limit, the vacuum expectation value of 
$\langle A_1(x^{\mu},y) \rangle$ becomes
\BQ
\langle A_1(x^{\mu},y) \rangle\Big|_{R = \infty} =
\frac{\sqrt{2}\Lambda}{g}\tanh\left(\frac{\Lambda}{\sqrt{2}}
(y-y_0) \right).
\label{kink}
\EQ
This is a single kink solution, which is just one of 
the topologically stable defects discussed in the paper \cite{DVALI}.
A key difference from the Dvali-Shifman approach is that the 
topologically stable solution (\ref{kink}) has been chosen as the
{\it vacuum} configuration in our model (but not chosen by hand).
This could be an advantage of our approach.
\par
We hope that our mechanism might shed new light on supersymmetry breaking. 
It would be of great importance to construct 
phenomenologically realistic supersymmetric models based on our approach.
\vskip0.3truein
\centerline{{\it ACKNOWLEDGMENTS}}
We would like to thank to H. Hatanaka and C. S. Lim
for useful discussions.
K.T. would like to thank the I.N.F.N, Sezione di Pisa for hospitality.
\newpage


\begin{thebibliography}{99}
\bibitem{LSC}
{I. Antoniadis, S. Dimopoulos and G. Dvali, 
Nucl. Phys. {\bf B516} (1998) 70 (hep-ph/9710204); \\
N. Arkani-Hamed, S. Dimopoulos and G. Dvali,
Phys. Lett. {\bf B429} (1998) 263 (hep-ph/9803315); \\
K. R. Dienes, E. Dudas and T. Gherghetta,
Phys. Lett. {\bf B436} (1998) 55 (hep-ph/9803466); \\
I. Antoniadis, S. Dimopoulos, A. Pomarol and M. Quiros,
hep-ph/9810410;\\
A. Delgado, A. Pomarol and M. Quiros, hep-ph/9812489.}
\bibitem{YANA}
{K. Izawa and T. Yanagida, Prog. Theor. Phys. {\bf 95} (1996) 829
(hep-th/9602180); \\
K. Intriligator and S. Thomas, Nucl. Phys. {\bf B473} (1996) 121
(hep-th/9603158).}
\bibitem{SAKA}
{M. Sakamoto, M. Tachibana and K. Takenaga, preprint 
KOBE-TH-99-01 (1999), IFUP-TH 6/99, (hep-th/9902069).}
\bibitem{OR}
{L. O'Raifeartaigh, Nucl. Phys. {\bf B96} (1975) 331.}
\bibitem{FER}
{S. Ferrara, L. Girardello and F. Palumbo, Phys. Rev. 
{\bf D20} (1979) 403.}
\bibitem{SAKA2}
{M. Sakamoto, M. Tachibana and K. Takenaga, in preparation.}
\bibitem{SCH}
{J. Scherk and J. H. Schwarz, Phys. Lett. {\bf B82} (1979) 60; \\
P. Fayet, Phys. Lett. {\bf B159} (1985) 121;
Nucl. Phys. {\bf B263} (1986) 87; \\
K. Takenaga, Phys. Lett. {\bf B425} (1998) 114 (hep-th/9710058); 
Phys. Rev. {\bf D58} (1998) 026004 (hep-th/9801075).}
\bibitem{DVALI}
{G. Dvali and M. Shifman, Nucl. Phys. {\bf B504} (1997) 127 
(hep-th/9611213).}
\end{thebibliography}
\end{document}